\documentclass[12pt]{article}
\def \1{\'{\i}}

\def \n{\noindent}
\def \&{&=&}
\def \t{\paragraph{$\bullet$}}
\def \nn{\nonumber}
\def \m{\medskip}
\newcommand{\be}{\begin{equation}}
\newcommand{\ee}{\end{equation}}
\newcommand{\beq}{\begin{eqnarray}}
\newcommand{\eeq}{\end{eqnarray}} 
\newcommand{\ba}{\begin{array}}
\newcommand{\ea}{\end{array}}  

\begin{document}
\setcounter{equation}{0}
\setcounter{section}{0}
\title{\Large \bf Non-classical symmetries and the singular manifold method: A further two
examples}
\author{ \bf P.G.
Est\'evez\footnote{e-mail: pilar@sonia.usal.es } and P.R. Gordoa  \\ {\small
\bf Area de F\1sica Te\'orica}\\ {\small \bf Facultad de F\1sica}\\ {\small \bf Universidad de
Salamanca}\\  {\small \bf 37008 Salamanca. Spain}\\} 
\maketitle
 
\begin{abstract} 

This paper discusses two equations with the conditional Painlev\'e property. The
usefulness of the singular manifold method as a tool for determining the non-classical symmetries
that reduce the equations to ordinary differential equations with the Painlev\'e property is
confirmed once more.  The examples considered in this paper are particularly interesting because
they have recently been
 proposed by other authors as counterexamples of the conjecture made by the authors that the
singular manifold method allows us to identify non-classical symmetries. We 
demonstrate here that the conjecture still holds for these two cases as well. A detailed study of
the   way of solving this apparent contradiction is offered.
\end{abstract} \vspace*{0.3in}

{\bf PACS Numbers 02.30 and 03.40K}
\newpage
\section{Introduction}
\setcounter{equation}{0}
In  1995 \cite{EG95} the present authors developed a method for identifying the non-classical
symmetries of partial differential equations (PDEs), using the Painlev\'e analysis as a tool 
\cite{WTC} and, more precisely, the Singular Manifold Method (SMM) based on the Painlev\'e
property \cite{P}, \cite{Weiss}. This paper was the continuation of two previous papers
\cite{E92}, \cite{E94} by one of us. In it, we studied six different PDEs. Four of them
were equations with the PP while the other two  considered there  were equations with only
the conditional  PP. The results obtained for these equations can be summarized as the following 
conjecture: {\it``The SMM allows one to identify the symmetries that reduce the original equation
to an ODE with the Painlev\'e property"}. Obviously, the combination of this statement with the ARS
conjecture 
\cite{Ars} means that for equations with the PP, the SMM should identify all the non-classical
symmetries. Nevertheless, for equations with the conditional PP, the SMM is only able to identify 
the symmetries for which the associated reduced ordinary differential equations (ODEs) are of
Painlev\'e type.

Recently, Tanriver and Roy Choudhury  \cite{TC} have applied our method to a family
of  Cahn-Hilliard equations. According to these authors, their results are apparently in
contradiction with ours because (according to them) for these equations the symmetries obtained 
using  the SMM are different from those obtained by the group theoretical non-classical method
\cite{OR1}.

If the conclusions of Tanriver and ChoudhuryÊ \cite{TC} were correct, the 
 Cahn-Hilliard equations would be  a counterexample that would cast some doubt on the
correctness of our  conjecture  \cite{EG95}.

In the following sections we shall prove that 
 \cite{TC} is incomplete and, consequently, that the  conclusions of those authors 
are flawed.
When the exercise is  done correctly, the results show that the  Cahn-Hilliard equations
are a further two good examples to be added to the list reported in
\cite{EG95}.

\section{Cahn-Hilliard equation for $m=1$ and one spatial variable}
\setcounter{equation}{0}
	This equation can be written as \cite{TC}:
\be u_t+\left (ku_{xx}-{u^2\over 2}\right)_{xx}=0\ee

\subsection{Non-classical method}
 The infinitesimal form of the Lie transformation of a PDE with two independent variables $x$ and
$t$ can be written as:
\beq \nn x' &=& x + \varepsilon\xi(x,t,u) + O(\varepsilon^2) ,  \\  \nn t' &=& t +
\varepsilon\tau(x,t,u) + O(\varepsilon^2) ,\\   u' &=& u +
\varepsilon\eta(x,t,u) + O(\varepsilon^2) , \eeq
\m
\n such that the associated Lie algebra contains vector fields of  the form
\be v = \xi {\partial\over\partial_x} +  \tau {\partial\over\partial_t} +  \eta
{\partial\over\partial_u} .
\ee
\m The non-classical method \cite{Ames}, \cite{BC}, \cite{OR1}, \cite{OR2}, \cite{FZ}
 requires that the symmetries should obey the invariant surface condition,
\be
\xi(x,t,u) u_x + \tau(x,t,u) u_t = \eta(x,t,u) 
\ee
\m 
\n associated with the vector field $v$.

The algorithmic method used to determine the equations to be satisfied by the infinitesimals 
$\xi$,
$\eta$ and $\tau$ is well known \cite{C1}, \cite{LW}, \cite{CW}. Nevertheless, as
 was mentioned several times in  
  \cite{EG95},  the non-classical method requires that the symmetries with 
$\tau=0$ should be determined separately from those with 
$\tau\neq 0$  \cite{C2}, \cite{C3}. Furthermore, there is no restriction in the use of the
normalization
$\xi=1$ when $\tau=0$. In the same way, 
  $\tau$ could be normalized to  $1$ when
$\tau\neq 0$ \cite{C2}, \cite{C3}. 
\subsubsection{Symmetries with $\tau=0$}

In this  case, we can choose 
$\xi=1$ without restriction, which means that the invariant surface condition is $\eta=u_x$.

The equation for $\eta$ is:
\beq \nn   &&  k\eta_{xxxx}+4k\eta\eta_{xxxu}+6k\eta^2\eta_{xxuu}+4k\eta^3\eta_{xuuu}
+k\eta^4\eta_{uuuu} +6k\eta_x\eta_{xxu}\\ 
\nn &&+6k\eta\eta_u\eta_{xxu} +12k\eta^2\eta_u\eta_{xuu} +12k
\eta\eta_x\eta_{xuu}  +6k\eta^2\eta_x\eta_{uuu}   +6k\eta^3\eta_u\eta_{uuu}    \\
\nn   &&+ 8k\eta\eta_{xu}^2+4k\eta_{xu}\eta_{xx}+
 4k\eta^3\eta_{uu}^2+3k\eta_{uu}\eta_x^2 
 +7k\eta^2\eta_u^2\eta_{uu}+4k\eta\eta_u^2\eta_{xu}\\
\nn && +12k\eta^2\eta_{xu}\eta_{uu}
+4k\eta_u\eta_x\eta_{xu}+10k\eta\eta_x\eta_u\eta_{uu}
+4k\eta\eta_{uu}\eta_{xx}\\ 
&&-u\eta^2\eta_{uu}-2u\eta\eta_{xu}-u\eta_{xx}-2\eta^2\eta_u-3\eta\eta_x+\eta_t=0
\eeq
This equation was obtained by using the {\tt symmgrp.max} MACSYMA
package 
\cite{CW}. The evident complexity of this equation  could be the reason why some
authors \cite{TC} have neglected these  symmetries. This complexity appears for many  $\tau=0$
symmetries \cite{C2}. Nevertheless, as  will be seen later on, one of the advantages of the SMM is
that it provides non-trivial solutions for (2.5).

\subsubsection{Symmetries with $\tau\neq 0$}

Calculation of these symmetries  \cite{CW}, \cite{TC} affords: 

\beq \nn \tau &=& 4\alpha t+\gamma \\  \nn \xi&=& \alpha x+\beta \\
\eta&=& -2 \alpha u \eeq

It is not difficult (see appendix) to check that the reduced ODEs associated with symmetries 
 (2.6) have the PP only in the following case

\medskip

\be \alpha=0, \qquad \beta=0\qquad\qquad \Longrightarrow \qquad\qquad
 \tau= 1, \qquad  \xi=0,\qquad   \eta=0\ee
where $\gamma$ has been normalized to $1$

\subsection{Singular manifold method}

Equation  (2.1) does not have the PP. However, it is possible to use the SMM to determine
particular solutions of (2.1) singlevalued on the initial conditions. For such solutions, the
equation has the conditional PP. To apply the SMM
\cite{Weiss},
\cite{WTC} we should look for solutions of (2.1) in the following form:

\be u=\sum_{j=0}^{\alpha}u_j\phi^{j-\alpha}\ee
where $\alpha$ and $u_0$ are respectively the leading  index and the leading term and $\phi$ is the
singular manifold that allows us to obtain truncated solutions such as in (2.8). Substitution of 
(2.8) in (2.1) provides two different expansions that depend on whether  the singular manifold
is  characteristic  ($\phi_x=0$) or not ($\phi_x\neq 0$). We shall explore both cases separately.
\subsubsection{Non-characteristic manifold}
If  $\phi_x\neq 0$, the expansion  (2.8) is \cite {TC}:
\be u'= u-12k\left({\phi_x\over \phi}\right)_x\ee
where $u$ is a solution of  (2.1) that could be expressed in terms of the singular manifold as:
\be u=4ks+3kv^2\ee
with $v$, $s$ and $w$ defined as:
\beq \nn v&=&{\phi_{xx}\over \phi_x}\\ \nn s&=&v_x-{v^2\over 2}\\ w&=&{\phi_t\over \phi_x}\eeq
It is worth noting  that  $w$ and $s$ are homographic invariants as oposed to 
$v$, which is not invariant under homographic transformations. 

The equations of the singular manifold are the equations satisfied by the homographic invariants 
$w$ and $s$  and  are:
\beq \nn w&=&0\\ s_x&=&s_t=0\eeq
The derivatives of  (2.10) can be written in terms of the singular manifold as:
\beq \nn u_x&=& v(u+2ks)\\ u_t&=& 0\eeq
where, according to \cite{EG95}, $v^2$ has been removed by using  (2.10). Substitution of
(2.13) in the invariant surface condition (2.4) is:
\be v(u+2ks)\xi=\eta \ee
The theory presented in  \cite{EG95} requires that the   invariant surface condition should only
depend on 
 {\bf  homographic invariants}. The infinitesimals must be determined in order to avoid the
presence of 
 $v$ in (2.14).
The only possibility of eliminating  the dependence on  $v$ of  the  invariant surface
condition is that $\xi=0$. This means that the only non-identically zero symmetry is:
\be  \tau=1,\qquad\qquad \xi=0 \qquad\qquad \eta=0 \ee
which is the non-classical symmetry (2.7)

\subsubsection{Characteristic manifold}
When  $\phi_x=0$, the truncated expansion  (2.8) is \cite{W2}:
$$u'=u-{1\over 6}(x+x_0)^2{\phi_t\over \phi}$$
where $u$ is a solution of  (2.1) whose expression in terms of the singular manifold is:
\be u={(x+x_0)^2\over 12}q(t)\ee
and where $q(t)$ has been defined as:
\be q(t)={\phi_{tt}\over \phi_t}\ee
Notice that for characteristic manifolds  \cite{EG95} the only homographic
invariant that we can construct is the schwartzian derivative with respect to $t$, defined as:
\be h=q_t-{q^2\over 2}\ee
in terms of which, the singular manifold equations   are:
\be h=0\ee
The derivatives of  (2.16) are:
\beq \nn u_x&=& {q\over 6}(x+x_0)\\ u_t&=&{q_t\over 12} (x+x_0)^2\eeq
(2.16) has to be used to remove $q$, or
$(x+x_0)^2$. The result is
\beq \nn u_x&=& {2u\over (x+x_0)}\\ u_t&=&{q u\over 2} \eeq

Since  $q$ is not  homographic invariant, we require that $\tau$ should be equal to zero in order to
avoid its presence in the invariant surface condition (2.4). The infinitesimals are in such a case:
\beq \nn \tau&=& 0\\ \nn  \xi&=&1 \\ \eta&=& {2u\over (x+x_0)}\eeq

It is easy to check that this symmetry satisfies  equation  (2.5) for the non-classical symmetries
with
$\tau=0$.

\subsection{Comparison of the non-classical method and SMM}
The SMM has allowed us to determine two different symmetries that are, respectively,   (2.15) and
(2.22). The former is the particular case of the non-classical symmetry (2.7) for which the
associated reduction leads to an  ODE with PP. The latter is a solution of  equation  (2.5) for
the non-classical symmetries with
$\tau=0$.

These results are in concordance with  \cite{EG95}. As we stated in the last example of this
reference  {\it `` [The symmetry identified by the SMM]    is the only one
in which the associated similarity reduction leads to an ODE of Painlev\'e type."}

\section{Cahn-Hilliard equation for $m=2$ and one spatial variable}
\setcounter{equation}{0}
	This equation can be written as \cite{TC}:
\be u_t+\left (ku_{xx}-{u^3\over 3}\right)_{xx}=0\ee

\subsection{Non-classical method}
In order to properly apply the non-classical method to equation (3.1), we 
  consider two different cases separately:
\subsubsection{Symmetries with  $\tau=0$}

If $\tau=0$, we can set $\xi=1$  with no  loss of generality. The resulting equation for
$\eta$ obtained using  \cite{CW} is:
\beq \nn
&& 4k\eta\eta_{xxxu}+k\eta_{xxxx}+k\eta^4\eta_{uuuu}+4k\eta^3\eta_{xuuu}+6k\eta^2\eta_{xxuu}+
6k\eta_x\eta_{xxu}+6k\eta^2\eta_x\eta_{uuu} \\
\nn &&+12k\eta\eta_x\eta_{xuu}+6k\eta^3\eta_u\eta_{uuu}+
12k\eta^2\eta_u\eta_{xuu}+6k\eta\eta_u\eta_{xxu}-u^2\eta_{xx}-u^2\eta^2\eta_{uu}-
2u^2\eta\eta_{xu} \\
\nn && +4k\eta_{xx}\eta_{xu}+3k\eta_x^2\eta_{uu}+4k\eta^3\eta_{uu}^2+
8k\eta\eta_{xu}^2+4k\eta\eta_{xx}\eta_{uu}+10k\eta\eta_x\eta_u\eta_{uu}+
4k\eta_x\eta_u\eta_{xu}\\
&& +12k\eta^2\eta_{xu}\eta_{uu} +7k\eta^2\eta_u^2\eta_{uu}+
4k\eta\eta_u^2\eta_{xu}-
6u\eta\eta_x-4u\eta^2\eta_u-2\eta^3+\eta_t=0
\eeq

\subsubsection{Symmetries with $\tau\neq 0$}

Solving the system of determining equations obtained using {\tt symmgrp.max} \cite{CW} yields
\cite{TC}:

\beq \nn \tau &=& 4\alpha t+\gamma\\  \nn \xi&=& \alpha x+\beta \\
\eta&=& - \alpha u\eeq

It can be shown (see appendix) that the reduced equations associated with the symmetries with
infinitesimal generators (3.3) only have the PP for the special choice of the parameters $\alpha=0$
and
$\beta=0$. In this case the infinitesimals are simply

\be \tau= 1, \qquad\qquad  \xi=0,\qquad\qquad   \eta=0\ee
where we have set $\gamma=1$ with no loss of generality.

\subsection{The singular manifold method}

 Equation (3.1) does not have  the PP as in the previous equation  (2.1). However, it
is possible, using the SMM, to search for particular solutions of (3.1) that are singlevalued
in the initial conditions. We therefore seek solutions of the
form \cite{Weiss}, \cite{TC}

\be u'=\sum_{j=0}^{\alpha}u_j\phi^{j-\alpha}\ee
The leading index is an integer only when the singular manifold
$\phi$ is non-characteristic ($\phi_x \neq0$), in which case  expansion (3.5) takes the form
\cite{TC} of:

\be u'= u+\sqrt{6k}\left({\phi_x\over \phi}\right)\ee
where $u$ is a solution of (3.1) that is expressed in terms of the singular
manifold $\phi$ as

\be u=-{\sqrt{6k}\over 2}v\ee
Furthermore, the singular manifold equations that relate $w$ and $s$
are

\beq \nn w&=&0\\ s&=&0\eeq
The next step is  to compute the derivatives of (3.7) in terms of
the singular manifold. The result is:

\beq \nn u_x&=& -{1\over \sqrt{6k}}u^2\\ u_t&=& 0\eeq
where  \cite{EG95} $v^2$  has been eliminated using (3.8).
Substitution of (3.9) in the invariant surface condition gives:

\be -{1\over \sqrt{6k}}u^2\xi=\eta \ee
According to equation (3.10) we must consider two cases: 

\medskip

\n $\bullet$ $\xi=0$

In this case $\eta=0$ and the only nontrivial symmetry we obtain is

\be \tau= 1, \qquad\qquad  \xi=0,\qquad\qquad   \eta=0\ee
which corresponds to the non-classical symmetry (3.4).

\medskip

\n $\bullet$ $\xi=1$

In this case equation (3.10)  is the invariant surface condition
associated whith a symmetry with $\tau=0$. The infinitesimal generators then take
 the form:

\be \tau=0 , \qquad\qquad  \xi=1,\qquad\qquad   \eta=-{1\over \sqrt{6k}}u^2\ee

It is trivial to check that (3.12) satisfies  equation (3.2) for the non-classical
symmetries with $\tau=0$.

\section*{Conclusions}
\t Two equations with the conditional PP  have been considered. Here we show that both of them
satisfy the conjecture established in
\cite{EG95} to the effect that the SMM allows us to identify all the non-classical
symmetries that reduce the equation to an ODE with the PP.

\t Our results do not agree with those found recently  by
Tanriver and Chowdhury \cite{TC}. This is because,
although these authors have tried to follow the method discussed in \cite{EG95}, they fail to
consider some of the aspects that were clearly stated in this reference. Such 
omissions lead them to   wrong conclusions and can be listed as:

\medskip

1) They do not take into account in both examples that computing
 the non-classical symmetries of an equation requires the consideration of  two different
cases separately, namely $\tau=0$ and $\tau\neq 0$. Only symmetries with
$\tau\neq 0$ were evaluated in \cite{TC}. However, as we have
shown in this paper, some of the symmetries of the solutions found
using the singular manifold method are symmetries with $\tau=0$.

\medskip

2) On applying the singular manifold method, they do not  consider
the case in which the singular manifold is characteristic
($\phi_x=0$). Solutions evaluated on the basis   of
characteristic manifolds turn  out to be relevant for equation (2.1), just as was the case of 
 some of the examples analyzed in \cite{EG95}.

\medskip

3) The authors of reference \cite{TC} should submit  \cite{EG95} to careful scrutiny.  In the
introduction to \cite{EG95} the following explicit statement was made:

\medskip

{\it ``We  show how for PDE with
Painlev\'e property, these symmetries are precisely those obtained through
the non-classical method. ..... Finally, for
equations with the conditional  Painlev\'e property, the SMM allows one to
identify the symmetries that reduce the original equation to an ODE with
the Painlev\'e property."}

According to the last sentence, and since equations (2.1) and (3.1) have
the conditional PP, the second part of the sentence applies for both
of them. As  has been shown for these examples (as well as  was
shown for example 6 in \cite{EG95}), the non-classical symmetries that
cannot be recovered through the SMM are precisely those that reduce the equation
to ODEs that are not of Painlev\'e type.

\t We believe that we have shown here that both equations, (2.1) and
(3.1), have not been interpreted correctly in \cite{TC}. Careful analysis
shows that the procedure developed in \cite{EG95} merely provides
two more examples that confirm the relationship between the non-classical
method and the SMM.

\paragraph{ACKNOWLEDGEMENTS}:

We would like to thank Professor P. Clarkson for the possibility of using some computer facilities
at the University of Kent and   Professor J. M. Cerver\'o for his advice. 

 This research has been supported in part by the DGICYT  under contract
PB95-0947.

\appendix
\section*{Appendix}
\subsection*{1.) Non-classical reductions for  symmetries (2.6)}
From the non-classical infinitesimal generators (2.6) we obtain two 
different reductions:

\t If $\alpha\neq 0$
\beq \nn u&=&{\alpha^2\over (\alpha x+\beta)^2}F(z)\\
\nn z&=&{(\alpha x+\beta)^4\over a^3(4\alpha t+\gamma)}\eeq
where $F(z)$ must satisfy the ODE:

 $$4k[32z^4F_{zzzz}+80z^3F_{zzz}+30z^2F_{zz}-15zF_z+15F]$$
$$-8z^2FF_{zz}-8z^2F_z^2+10zFF_z-2z^2F_z-5F^2=0$$

\medskip

\n which is not of Painlev\'e type.

\t If $\alpha= 0$
\beq \nn u&=&F(z)\\
\nn z&=&\gamma x-\beta  t\eeq
where $F(z)$ is a solution of

 $$-\beta F_z +k\gamma^4F_{zzzz}-\gamma^2(FF_{zz}+F_z^2)=0$$

\medskip

\n It can be easily shown that this equation is not of Painlev\'e type
unless $\beta=0$.
\subsection*{2.) Non-classical reductions for symmetries (3.3)}
Similarly, we obtain two different reductions from the
symmetry (3.3):

\t If $\alpha\neq 0$
\beq \nn u&=&{\alpha\over (\alpha x+\beta)}F(z)\\
\nn z&=&{(\alpha x+\beta)^4\over a^3(4\alpha t+\gamma)}\eeq
where $F(z)$ must satisfy

 $$4k[64z^4F_{zzzz}+224z^3F_{zzz}+108z^2F_{zz}-6zF_z+6F]$$
$$-16z^2F^2F_{zz}+12zF^2F_2-z^2F_z-32z^2FF_z-4F^3
=0$$

\medskip

\n which is not of Painlev\'e type.

\t If $\alpha= 0$
\beq \nn u&=&F(z)\\
\nn z&=&\gamma x-\beta  t\eeq
where $F(z)$ is a solution of 

 $$-\beta F_z +k\gamma^4F_{zzzz}-\gamma^2(F^2F_{zz}+2FF_z^2)=0$$

\medskip

\n This equation is again of Painlev\'e type only when $\beta=0$.

\end{document}